# DENSITY-FUNCTIONAL THEORY FOR THE HUBBARD MODEL: NUMERICAL RESULTS FOR THE LUTTINGER LIQUID AND THE MOTT INSULATOR

K. Capelle

*Departamento de Química e Física Molecular*
*Instituto de Química de São Carlos, Universidade de São Paulo,*
*Caixa Postal 780, São Carlos, 13560-970 SP, Brazil*

N. A. Lima, M. F. Silva, and L. N. Oliveira

*Departamento de Física e Informática*
*Instituto de Física de São Carlos, Universidade de São Paulo,*
*Caixa Postal 369, 13560-970 São Carlos, SP, Brazil*

**Abstract**     We construct and apply an exchange-correlation functional for the one-dimensional Hubbard model. This functional has built into it the Luttinger-liquid and Mott-insulator correlations, present in the Hubbard model, in the same way in which the usual *ab initio* local-density approximation (LDA) has built into it the Fermi-liquid correlations present in the electron gas. An accurate expression for the exchange-correlation energy of the homogeneous Hubbard model, based on the Bethe Ansatz (BA), is given and the resulting LDA functional is applied to a variety of inhomogeneous Hubbard models. These include finite-size Hubbard chains and rings, various types of impurities in the Hubbard model, spin-density waves, and Mott insulators. For small systems, for which numerically exact diagonalization is feasible, we compare the results obtained from our BA-LDA with the exact ones, finding very satisfactory agreement. In the opposite limit, large and complex systems, the BA-LDA allows to investigate systems and parameter regimes that are inaccessible by traditional methods.

## 1.    The Hubbard model and density-functional theory

The Hubbard model is one of the most venerable models of many-body physics. Originally it was proposed as a simplified description of magnetism in transition metals [1, 2]. This required the model to be formulated on a three-dimensional lattice. Interest in the two-dimensional Hubbard model is more recent, and largely due to Anderson's suggestion that it contains the correct minimal requirements needed for describing the cupper-oxide planes in cuprate high-temperature superconductors [3]. In one dimension the Hubbard model has attracted interest mainly because it presents a fascinating phase diagram, including Luttinger liquid and Mott insulating phases which are at the center of much recent work on strongly correlated systems (see, e.g., Refs. [4, 5, 6, 7, 8] for reviews). Independently of the issue of strong correlations, the one-dimensional Hubbard model (1DHM) has aquired additional significance with the recent experimental confirmation of Luttinger liquid behaviour in quasi one-dimensional systems such as carbon nanotubes [9, 10, 11], and quantum wires [12, 13]; systems that offer great potential for applications in the field of nanotechnology.

In its simplest form the 1DHM reads

$$\hat{H} = -t \sum_{<ij>\sigma} c_{i\sigma}^\dagger c_{j\sigma} + U \sum_i c_{i\uparrow}^\dagger c_{i\uparrow} c_{i\downarrow}^\dagger c_{i\downarrow}, \qquad (1)$$

where $t$ is the hopping matrix element, $U$ the on-site interaction, and the sum on $i, j$ is restricted to nearest neighbours. Motivated both by the traditional (strong correlations) and more recent (nanotechnology) interest in the 1DHM, we have recently embarked on a density-functional analysis of this model and some of its extensions [14, 15].

In principle, one could think of several ways in which density-functional theory (DFT) and the Hubbard model can be brought together. One attractive possibility is to use *ab initio* DFT in order to calculate the parameters $t$ and $U$ from first principles. Examples of this approach are Refs. [16, 17]. Alternatively, one can try to incorporate the physics of the Hubbard-model into approximate density functionals of *ab initio* DFT. An example of this line of thought is the so-called LDA+U method, in which a Hubbard $U$ is introduced into the *ab inito* LDA functional [18, 19]. Still another possibility is to use the Hubbard model as a laboratory in which formal questions of DFT (such as the meaning of the Kohn-Sham eigenvalues or the band-gap problem) can be studied. This kind of study was pioneered by Gunnarsson and Schönhammer [20, 21, 22]. Finally, one can consider the Hubbard Hamiltonian (1) as a many-body problem in its own right (and a quite difficult one at that), to which DFT can be applied as a calculational tool [14]. In this paper we are concerned with the latter two possibilities.

A prerequisite for applying DFT to the Hubbard Hamiltonian is, of course, a reformulation of the Hohenberg-Kohn theorem and the Kohn-Sham equations, which were originally formulated for the *ab inito* Hamiltonian and not for model Hamiltonians. This reformulation was accomplished by Gunnarsson and Schönhammer [20, 21], who set up so-called *site-occupation DFT*, in which the occupation number $n_i$ of site $i$ plays the same role as the particle density $n(\mathbf{r})$ does in *ab initio* DFT. Recently we have constructed an explicit and simple

LDA-like density functional for the 1DHM, which can be used in conjunction with the Gunnarsson-Schönhammer form of the Kohn-Sham equations [14, 15]. Here, we explain this construction in detail, and present some numerical results for a variety of inhomogeneous Hubbard models.

Both the philosophy and the technical details of the construction of the 1DHM functional are very similar to those of the *ab initio* LDA. To appreciate this similarity, let us briefly recall the construction of the latter. First, one considers a homogeneous interacting electron gas (a charged Fermi liquid) and calculates its total energy as a function of the particle density. This calculation was first performed using perturbation theory [23, 24], and more recently with Quantum-Monte Carlo (QMC) [25]. Next, one subtracts the noninteracting kinetic energy and the Hartree energy to extract the exchange-correlation ($xc$) energy. The result of these two steps is a numerically defined $xc$ functional. For practical applications one needs a sufficiently simple and simultaneously reasonably accurate parametrization of the numerical data. Such parametrizations are constructed taking into account known exact results, such as high-density and low-density limits or scaling properties [23, 24, 26, 27, 28]. Finally, the resulting analytical expression for the per-volume $xc$ energy of the homogeneous electron gas, $e_{xc}(n)$, is used locally to approximate the $xc$ energy of the inhomogeneous real system,

$$E_{xc}^{LDA}[n] = \int d^3r \, e_{xc}^{hom}(n)|_{n \to n(\mathbf{r})}. \qquad (2)$$

In this way one transfers the correlations present in one's reference system, the charged Fermi liquid, into the DFT description of the real inhomogeneous system under study.

Let us now consider the one-dimensional Hubbard model. Here the homogeneous reference system is given by Eq. (1). This model is known [4, 29] to describe a charged Luttinger liquid (except for the case of a half-filled band, where it is a Mott insulator). In comparison with the *ab initio* case, where one needs perturbation theory or QMC to calculate the energy of the reference system, the situation for the 1DHM is rather more favorable, since an *exact* solution is available. This solution is obtained by means of the Bethe Ansatz (BA) [5, 30, 31]. The BA results in a set of coupled integral equations, which must still be solved numerically and simplify only in the $U \to \infty$ and $U \to 0$ limits. In these limits, however, one can extract analytical formulae for the total energy as a function of the site occupation numbers $n_i$ and $U$. We use these formulae, together with a similar expression valid at any $U$ for precisely half filling (all $n_i = 1$), to construct a simple expression for the per-site total energy of the homogeneous 1DHM. The following steps are then precisely as in the *ab inito* case: we subtract kinetic and Hartree energies and use the result to locally approximate the $xc$ energy of inhomogeneous Hubbard models,[1]

$$E_{xc}^{LDA}[n] = \sum_i e_{xc}^{hom}(n)|_{n \to n_i}. \tag{3}$$

Our expression for $e_{xc}(n)$ is derived in the next section, and numerical results obtained from it are shown in the remainder of this paper. Just as in the *ab initio* case, DFT allows us to study inhomogeneous interacting systems by diagonalizing the Hamiltonian of a noninteracting (Kohn-Sham) system. Interesting inhomogeneities include boundaries, surfaces, charge- and spin-density waves, dimerization, impurities, etc. Most of these can be modeled by adding an additional on-site potential

$$\hat{V} = \sum_{i,\sigma} v_i c_{i\sigma}^\dagger c_{i\sigma} \tag{4}$$

to the homogeneous Hamiltonian (1), others require more complicated extra terms. Some examples are given below.

Before we enter these details, however, we briefly mention two alternative LDA-type functionals that have been applied to the 1DHM. One, below called the 'pseudo-LDA', is a discretized form of the three-dimensional *ab initio* LDA [20, 21]. This functional has not performed very well numerically [15, 32] and has been criticized also on the fundamental grounds that the three-dimensional electron gas, on which the *ab initio* LDA is based, is not the correct reference system for the one-dimensional Hubbard model [15, 33, 34]. The other is a numerical Bethe-Ansatz based LDA [22]. This approach is conceptually similar to ours, but has not resulted in an explicit density functional. Instead, it relies on numerical solution of the Lieb-Wu integral equations for the ground-state energy; and subsequent numerical differentiation in order to obtain the corresponding potential.

## 2. Exchange-correlation energy of the Hubbard model

From the exact Bethe-Ansatz solution to the homogeneous 1DHM [31] one can extract analytical expressions for the total ground state energy in several important limiting cases. For infinitely strong interactions ($U \to \infty$) and a less than half-filled band ($n < 1$), e.g., one has [5]

$$e(n, t, U \to \infty) = -\frac{2t}{\pi} \sin(\pi n) \tag{5}$$

---

[1] Due to the particular form of the on-site interaction only electrons with opposite spins interact in the Hubbard model, so that there is no exchange energy in the quantum-chemical sense. Nevertheless, we use the expressions 'exchange-correlation energy' and 'exchange-correlation functional', in order to emphasize that these are the direct Hubbard counterparts to the corresponding *ab initio* concepts.

where $n = N/L$ is the band filling (a constant in the homogeneous case), $N$ the number of electrons and $L$ the number of lattice sites. In the absence of interactions ($U = 0$) one straightforwardly obtains

$$e(n, t, U = 0) = -\frac{4t}{\pi} \sin\left(\frac{\pi}{2}n\right). \tag{6}$$

Finally, for an exactly half-filled band ($n = 1$) and any interaction $U$ [5]

$$e(n = 1, t, U) = -4t \int_0^\infty dx \frac{J_0(x) J_1(x)}{x(1 + \exp(xU/2t))}, \tag{7}$$

where $J_0$ and $J_1$ are zero and first order Bessel functions.

We now employ these three results to set up an interpolation formula for intermediate values of $U$ and $n$. Motivated by the similarity of the limiting expressions (5) and (6) we adopt the functional form

$$e(n, t, U) = -\frac{2t\beta(U/t)}{\pi} \sin\left(\frac{\pi}{\beta(U/t)}n\right), \tag{8}$$

where $\beta$ is a function of the ratio $U/t$. Clearly, for $U \to \infty$ one must have $\beta = 1$, while for $U = 0$ one must recover $\beta = 2$. To fix $\beta$ for intermediate values of $U/t$ we employ Eq. (7) and determine $\beta$ from requiring that

$$-\frac{2t\beta(U/t)}{\pi} \sin\left(\frac{\pi}{\beta(U/t)}\right) = -4t \int_0^\infty dx \frac{J_0(x) J_1(x)}{x(1 + \exp(xU/2t))}, \tag{9}$$

which guarantees the correct result at half filling ($n = 1$). This requirement in fact determines $\beta$ for all values of $U$, including the limiting cases $U \to \infty$ and $U = 0$, since in these two limits one can calculate the integral analytically and indeed recovers $\beta = 1$ and $\beta = 2$, respectively.

Note that the integral appearing in Eq. (9) does not lead to computational complications: for any given value of $U$ (i.e., for any fixed Hamiltonian) the right-hand side is just a number (the integral in it is easily calculated numerically and converges rapidly). Having determined this number, Eq. (9) is merely a transcendental equation for $\beta$, which can be solved by standard methods and has exactly one solution in the physical interval ($U = 0, U \to \infty$), that is, in the interval ($\beta = 1, \beta = 2$). This entire calculation takes place outside the self-consistency cycle of DFT.

The interpolation formula (8) with (9) is already our final result for $n \leq 1$, i.e., up to half filling. For a more than half-filled band a particle-hole transformation [31, 5] can be used to express the energy in terms of that for a less than half-filled band,

$$e(n > 1, t, U) = e(2 - n, t, U) + U(n - 1), \tag{10}$$

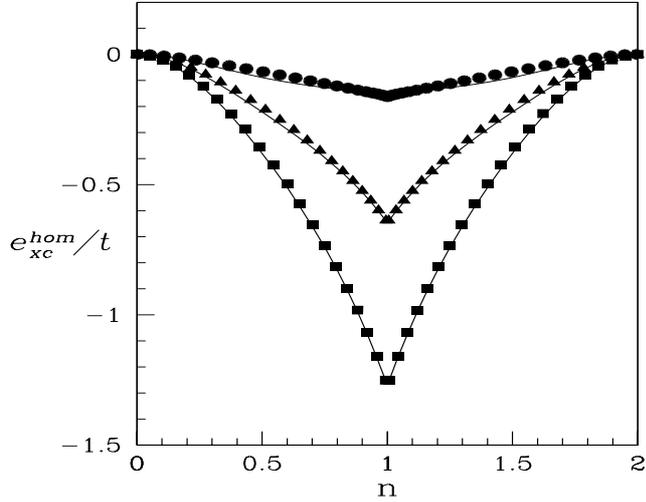

*Figure 1.* Exchange-correlation energy per site of the homogeneous infinite 1DHM as obtained by numerically solving the Lieb-Wu integral equations resulting from the Bethe Ansatz. Circles: $U = 3$, triangles: $U = 6$, squares: $U = 9$. The full lines are obtained from our expression (8) with (9) and (10). The band filling $n$ ranges from $n = 0$ (empty band) over $n = 1$ (half-filled band) to $n = 2$ (filled band). The form of the curves reflects particle-hole symmetry, and the kinks at $n = 1$ signal the Mott metal-insulator transition.

where $e(2 - n, t, U)$ is the energy for a less than half-filled band, given above. This completes our interpolation of the ground-state energy of the 1DHM. The quality of the expression obtained is illustrated in Fig. 1, in which we compare the ground-state energy calculated from Eqs. (8) with (9) and (10) with the one obtained from numerically solving the Lieb-Wu integral equations following from the Bethe Ansatz.

In Fig. 2 we show how our expression interpolates between the $U = 0$ and $U \to \infty$ limits. For comparison purpose we have included in this figure also two curves representing the corrections to the $U \to \infty$ limit up to order $1/U$ and $1/U^3$, respectively [35]. Our interpolation is by construction exact at $U = 0$ and $U \to \infty$. At large $U$ it is very similar to the asymptotic expansions, but unlike these it recovers the correct $U = 0$ limit.

Note that here our procedure slightly deviates from the one common in *ab initio* DFT: we have based our expression for $e(n, t, U)$ entirely on the three exact limiting cases, and used the numerical Bethe Ansatz data only for comparison purposes. In the *ab initio* case the LDA functional is based directly on a parametrization of the numerical QMC data, and the exactly known limits serve only as constraints [26, 27, 28]. We could adopt a similar procedure here by introducing a number of free parameters in the functional and fit these to the numerical data. This would clearly provide an even better parametrization of the total energy than the above interpolation, but for our present purposes that interpolation is sufficiently accurate.

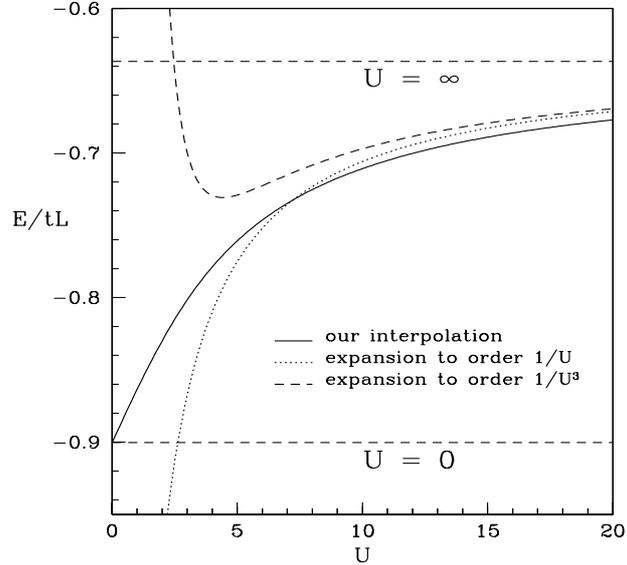

*Figure 2.* Full curve: total energy of the homogeneous infinite 1DHM with $n = 0.5$, as calculated from our expression (8) with (9) and (10), as a function of interaction strength $U$. The two dashed horizontal lines denote the limits $U = 0$ and $U \to \infty$, respectively. The dotted and dashed curves are the analytically known [35] correction to the $U \to \infty$ result to order $1/U$ and $1/U^3$, respectively.

Given an expression for the total energy of the homogeneous 1DHM one can use it in either of two ways. First, many interesting observables can be expressed in terms of this energy. If one uses the Bethe Ansatz directly to calculate these, one obtains complicated expressions that must be evaluated numerically and simplify only in the $U = 0$ and $U \to \infty$ limits. On the other hand, a simple approximate expression for $e(n,t,U)$ can be used to obtain simple analytical results also between these limits. An example is the Mott gap that opens at half filling in the homogeneous 1DHM. This gap can be calculated as $\Delta = I - A$, where $I$ and $A$ are ionization energy and electron affinity, respectively. Both of these quantities can be calculated as differences of total ground-state energies. We have recently used our expression for this energy to obtain the following approximate expression for the Mott gap in the thermodynamic limit [15]

$$\Delta(U) = U + 4t \cos\left(\frac{\pi}{\beta(U/t)}\right). \tag{11}$$

This expression can be shown [15] to yield the correct results in the limits $U \to \infty$ and $U = 0$. In between these limits its accuracy is comparable with that of the asymptotic expansion of $\Delta$ to order $1/U$ [36]. More details are given in Sec. 3.4 and Ref. [15]. Many other quantities can be treated in the same way. Research along these lines is currently under way in our group.

A second possible use one can make of the expression for $e(n, t, U)$ is to employ it as an input for constructing an LDA-type functional. To this end one must subtract the per-site noninteracting kinetic energy $t_s$ and Hartree energy $e_H$ from $e(n, t, U)$. We define the Hartree energy in general as

$$E_H[n, U] = \frac{U}{4} \sum_i n_i^2, \tag{12}$$

where $n_i = \langle c_{i\uparrow}^\dagger c_{i\uparrow} + c_{i\downarrow}^\dagger c_{i\downarrow} \rangle$ is the local occupation number. (Other definitions are possible, but this one is convenient for our purposes. As long as one uses a consistent expression for the Hartree term in the definition of the $xc$ functional and in the KS equations all choices are equivalent on the exact level.) For a homogeneous system our choice implies $e_H(n, U) = U n^2/4$. The noninteracting kinetic energy is simply given by $e(n, t, U = 0)$, since for $U = v_i = 0$ the Hamiltonian of the 1DHM contains only the kinetic energy term. Our functional then becomes

$$E_{xc}^{LDA}[n, t, U] = \sum_i e_{xc}(n, t, U)|_{n \to n_i}, \tag{13}$$

where

$$e_{xc}(n, t, U) = e(n, t, U) - e(n, t, 0) - e_H(n, U), \tag{14}$$

$e(n, t, U)$ is our expression (8) with (9) for $n_i \leq 1$, and use of (10) is implied for $n_i > 1$.

From the point of view of formal DFT it is worthwhile to point out that this functional explicitly depends on the interaction $U$ and the hopping parameter $t$. In fact, this should not come as a surprise: even the *ab initio* $xc$ functional depends on the parameter determining the interaction strength and the coefficients in the kinetic energy operator. The only difference is that in the *ab initio* case these are usually fixed to be $e^2$ for the interaction and $\hbar^2/2m$ for the kinetic energy, and one does not bother to specify the dependence on $e^2$ and $m$ in the functionals.

The appearance of these parameters in the density functional is a necessary consequence of the fact that the Hohenberg-Kohn theorem asserts universality of the functional with respect to the external potential $\hat{V}$, but not with respect to the interaction law or the form of the kinetic energy. Even in *ab initio* DFT one requires new functionals when one considers, e.g., phonon-induced electron-electron interactions (such as in DFT for superconductors [37]), or the Dirac kinetic energy (such as in relativistic DFT [38]). In the present, Hubbard, case the dependence of the functional on $U$ and $t$ is strongly constrained: if energies are scaled by $t$, then all properties of the Hubbard Hamiltonian, including the $xc$ functional, depend only on the ratio $U/t$. Our interpolation, as constructed above, respects this condition. Below we thus follow the universally adopted convention to take $t$ as our unit of energy, and do not include it explicitly among the parameters in the functional.

A final remark on our functional is that it is, of course, far from optimal, and offers much opportunity for improvement. Among other things one could consider to (i) use spin-resolved densities instead of the charge density, i.e., construct an LSDA instead of an LDA, (ii) parametrize the integral (7) to obtain a closed expression for $\beta(U)$, (iii) use scaling conditions to separate exchange and correlation contributions to $e_{xc}$, (iv) develop a parametrization directly for the numerical data obtained from the Lieb-Wu integral equations, (v) apply self-interaction corrections, (vi) extend the interpolation to negative values of $U$ (interesting in connection with purely electronic superconductivity), etc. Several of these projects are currently under study in our group.

## 3. Applications

### 3.1 Luttinger liquids

'Luttinger liquid' is the name usually given to one-dimensional Fermi liquids [4]. One-dimensional metals, for which the Luttinger liquid is the unifying paradigm, behave in many ways so differently from their higher dimensional counterparts (e.g., they do not have low-lying fermionic quasi particles) that a whole new set of concepts and a very specific terminology has been developed to deal with them [4, 5, 29]. Although one-dimensional metals may appear as a theoretical curiosity, there are many systems, such as carbon nanotubes [9, 10, 11], quantum wires [12, 13], edge states in the fractional quantum Hall effect [39, 40], and quasi one-dimensional organic [41, 42] and inorganic [43, 44, 45] conductors, for which experiments indicate Luttinger liquid behaviour. The recent upsurge of interest in nanoscale physics has brought in particular quantum wires and carbon nanotubes into the focus of mainstream research, and as a consequence Luttinger liquid theory has aquired a quite unexpected relevance for device technology and related applications.

The Luttinger model (from which the universality class of one-dimensional metals derives its name) is perhaps the simplest model whose low-energy degrees of freedom are described by Luttinger-liquid phenomenology. For our purposes, however, it is more important that the homogeneous 1DHM for band-fillings $n \neq 1$ (i.e., off half filling) is also a Luttinger liquid. Our functional can be directly applied to this phase of the 1DHM, and used to extract a variety of observables. Here we just consider the ground-state energy; results for other quantities will be published separately.

In Fig. 3 we plot the total energy of a homogeneous 1DHM with $U = 6$ as a function of lattice size, both for open (Hubbard chains) and periodic (Hubbard rings) boundary conditions. For even $L$ we take $N = L/2$ (so that $n = N/L = 1/2$, corresponding to quarter filling). For odd $L$ we take $N = (L-1)/2$. BA-LDA results, obtained from self-consistent solution of the 1DHM Kohn-Sham equations with our functional, are compared with numerically exact ones. The exact (Lanczos) diagonalization was carried up to $L = 14$ on a small PC, taking into account the conservation of particle number and of the z-component of the spin, and advantage of sparse matrix techniques. With a supercomputer, and exploiting symmetries such as total spin, particle-hole symmetry, etc. one can perhaps double the maximum attainable size, but the exponentially increasing Hilbert space makes such calculations prohibitively expensive, and with today's computing technology there is no way of attaining, e.g., $L = 100$.

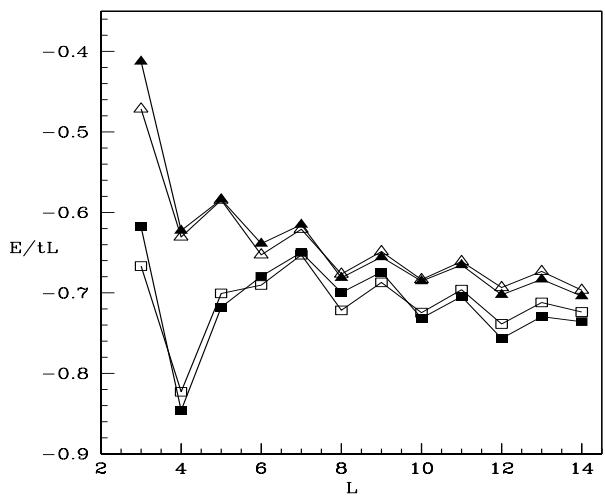

*Figure 3.* Total energy of a finite 1DHM in the Luttinger liquid phase for $U = 6$. Full triangles: BA-LDA results for open boundary conditions. Open triangles: exact results for open boundary conditions. Full squares: BA-LDA results for periodic boundary conditions. Open squares: exact results for periodic boundary conditions. The lines are only a guide for the eye.

The following conclusions can be drawn from these calculations: (i) Even for very small systems the BA-LDA is reliable. It faithfully reproduces the difference between both types of boundary condition and the even-odd oscillations as a function of the number of lattice sites, with an error of not more than a few percent. In view of the thermodynamic limit built into the reference system for the LDA, this good performance even for small systems is a welcome surprise. (ii) The difference between the exact and the LDA results for moderate $L$ is largely due to the fact that our expression (8) for the energy of the homogeneous reference system is only an interpolation between exact results, but not itself exact for all values of the parameters. A better parametrization (work on which is in progress) would presumably diminish the remaining differences between the exact and LDA curves. (iii) The difference between results obtained for open and for closed boundary conditions only becomes small when the system size exceeds the range accessible with exact diagonalization. The common tenet of solid-state physics 'boundary-conditions do not matter for bulk phenomena' thus is only true when the bulk is already too large for exact calculations to be viable. (iv) The computational effort for the BA-LDA is orders of magnitude lower than for the exact diagonalization, since one must only diagonalize a noninteracting Hamiltonian. In fact it is no problem at all to calculate the energy and other observables for hundreds of sites on a small PC.

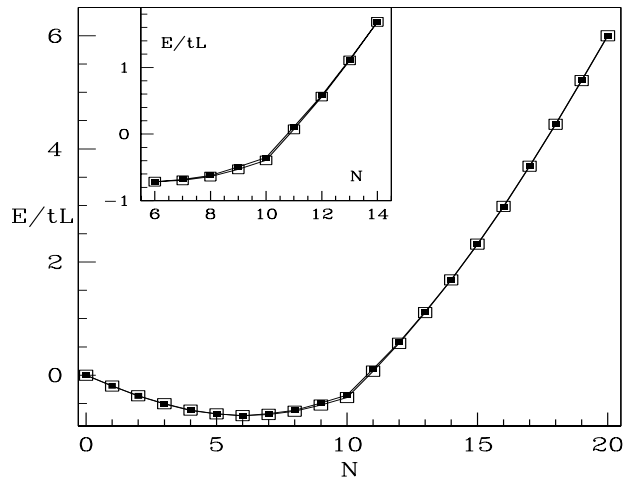

*Figure 4.* Full squares: total energy of a 1DHM with open boundary conditions, $U = 6$ and $L = 10$, calculated from the BA-LDA as a function of the total number of electrons. Open squares: exact results obtained by numerically diagonalizing the 1DHM. The inset is a zoom into the region near $N = L$, where the BA-LDA functional is discontinuous and the 1DHM undergoes its metal-insulator transition. The lines are a guide for the eye.

Density-matrix renormalization group (DMRG) [46] calculations are capable of attaining similar system sizes, and usually achieve much better accuracy. However, DMRG calculations are encounter difficulties for periodic boundary conditions, are hard to apply to inhomogeneous systems (see below), and become computationally expensive for $L$ in the hundreds or larger. BA-LDA, on the other hand, is less accurate, but does not suffer from either of these drawbacks. A BA-LDA calculation can thus provide useful complementary information to a DMRG one.

As another illustration of the BA-LDA applied to the Luttinger-liquid phase, we plot, in Fig. 4, the energy for a fixed system size $L = 10$ and vary the number of electrons $N$. Again, exact and BA-LDA results agree well. The abrupt change of slope at $N = L$ is a signal of the metal-insulator transition taking place at half filling.

## 3.2 Impurity models

In the examples of the previous section the inhomogeneity arose only from the finite size of the system, which was still homogeneous in the bulk. More interesting both from a fundamental and a practical point of view, are systems that are inhomogeneous also in the bulk. An example for such an inhomogeneity that has been much studied in the literature is that of an impurity in a Luttinger liquid. The effect of an impurity on a one-dimensional system is much more profound than on a three dimensional system, because for open boundary conditions the impurity effectively splits the system in two subsystems. Particles can get from one subsystem to the other only by passing through the impurity site. This is to be compared with the situation in a three-dimensional system, in which particles can circumvent the impurity site in many ways. Similarly, for periodic boundary conditions in one dimension there is no closed path that does not involve the impurity, while there are many in higher dimensions.

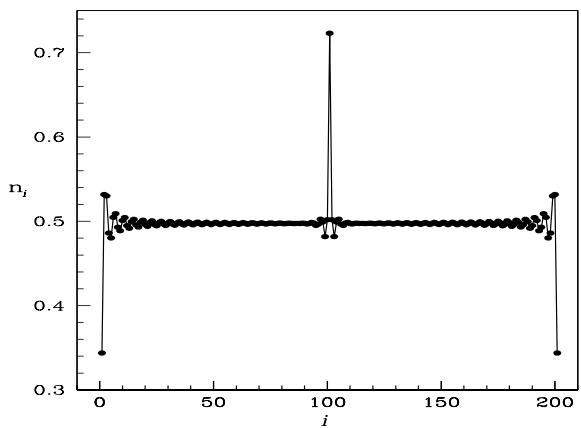

*Figure 5.* Density distribution for an impurity in a Luttinger liquid with $U = 6$, $n = 0.5$, and open boundary conditions. The impurity is described by an on-site potential of unit strength at the central site. The Friedel oscillations arising at the impurity are clearly visible, and comparable in size with those originating ate the surfaces. The lines are a guide for the eye.

One consequence of this different physics in one dimension is very pronounced Friedel oscillations arising around the impurity. Another is that the convergence to the thermodynamic limit is significantly slowed down by the presence of the impurity. We have discussed these issues in Ref. [14]. Here we provide, for illustration, a plot of the density distribution around the impurity site. Fig. 5 clearly displays the Friedel oscillations arising from the impurity and from the boundary (which in a finite system with open boundary conditions also acts as an impurity).

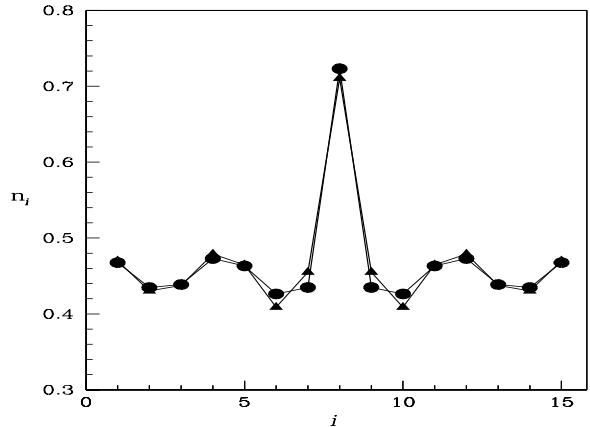

*Figure 6.* Density distribution of a 15-site 1DHM with an inpurity of strength $v_I = -1$ on the central site ($U = 5$, $n = 1$, periodic boundary conditions). Circles: exact result. Triangles: BA-LDA. The lines are a guide for the eye.

In Fig. 6 we consider a system with $L = 15$ sites, for which exact diagonalization is still possible, and compare the BA-LDA results for the density oscillations with the exact ones. For such small systems the LDA is not expected to do well (recall that it was, just as any other LDA, based on the thermodynamic limit and assumes slow spatial variation of the density). However, in spite of this caveat the LDA density distribution is seen to agree quantitatively with the numerically exact one.

For larger systems, such as that of Fig. 5, exact diagonalization becomes prohibitive. For not too large systems one can compare with DMRG, but even that method is numerically very expensive for impurities in the bulk. As a computationally less expensive alternative one can place the impurities at the boundaries. A systematic DMRG studies of such boundary fields in the 1DHM has been performed in Ref. [47]. For the BA-LDA approach the limits on system size are much less restraining than for other methods, and it makes little difference where one places the impurity. As an example for a calculation with boundary fields, we display here, in Fig. 7, the density at site $i = 1$ for the case in which impurities are located at the boundaries, i.e., at $i = 1$ and $i = L$. In this calculation we have choosen exactly the same parameters as in figure 4 of Ref. [47], so that one can directly compare both results. There are some data points missing for densities very close to $n = 1$. This is due to a convergence problem in the self-consistency cycle, and will be discussed in Sec. 3.4. Apart from this region the BA-LDA curves and the corresponding DMRG curves are in reasonable, but not yet optimal, agreement: the numerical values and global behaviour are very similar, but the local curvature of the BA-LDA results is visibly different from the DMRG one. Improvements on the functional (along the lines described at the end of Sec. 2 and of Sec. 3.4, respectively) are expected to further improve the agreement with the DMRG results.

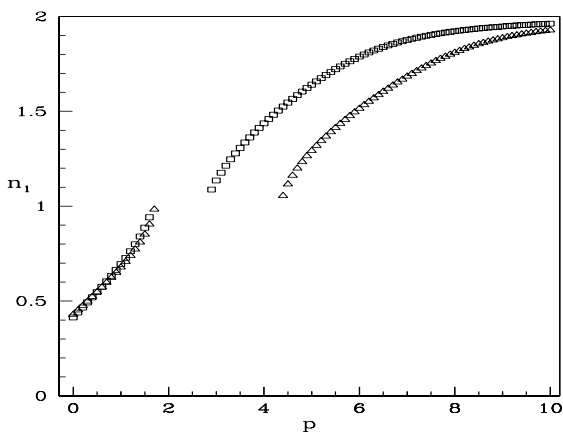

*Figure 7.* Density at site $i = 1$ for an impurity model in which the impurities of strength $p$ are located at the system's boundary. This plot, obtained with the BA-LDA, is to be compared with Fig. 4 of Ref. [47], obtained using density-matrix renormalization group for the same model. Upper curve (squares): $U = 4$, lower curve (triangles) $U = 6$. In both cases $n = 0.55$ and $L = 100$.

Quite independently of these details, the degree of agreement between the BA-LDA and the exact and DMRG results is remarkable, in view of the fact that the BA-LDA is orders of magnitude faster and less memory-consuming than these more precise methods. The BA-LDA may thus be a useful tool to explore parameter regimes (in particular in large and/or inhomogeneous systems) that are impossible to access with standard methods. An example in which this expanded range of accessible parameter space is immediately useful is the determination of anomalous exponents in Luttinger liquids. These exponents, which govern the asymptotic behaviour of density distributions, correlation functions, etc., are among the parameters that characterize a given Luttinger liquid, and their values and dependence on the interaction can only be extracted from Luttinger liquid theory through some fairly involved mathematics — if at all [4]. Numerical determination of such exponents is hampered by the fact that they become well defined only asymptotically, and systems that are large enough to allow the asymptotic regime to take over are exceedingly hard to treat with traditional methods. The Bethe-Ansatz LDA then provides an attractive alternative. We have given a first example for the determination of such exponents from BA-LDA density distributions in Ref. [14]. A more detailed study is under way.

### 3.3  Spin-density waves

We now turn to a case in which the inhomogeneity does not occur in real space, but in spin space. Specifically, we add the term

$$\hat{S} = S \sum_{\langle ij \rangle} \left( c_{i\uparrow}^\dagger c_{j\downarrow} + H.c. \right) \tag{15}$$

to the Hamiltonian (1). The *staggered field* $S$ couples spin up and spin down states. For $i = j$ the combination of creation and annihilation operators $c_{i\uparrow}^\dagger c_{i\downarrow} =: \hat{\rho}_{s,i}$ is related to the $x$ and $y$ components of the local spin magnetization via

$$\hat{m}_x(\mathbf{r}) = \mu_0 [\hat{\rho}_{s,i} + \hat{\rho}_{s,i}^\dagger] \tag{16}$$

$$\hat{m}_y(\mathbf{r}) = i\mu_0 [\hat{\rho}_{s,i} - \hat{\rho}_{s,i}^\dagger], \tag{17}$$

where $\mu_0$ is the Bohr magneton. A spin configuration that is not completely specified by the $z$-component of the full magnetization vector $\mathbf{m}$, but requires specification of the $x$ and $y$-components as well, is noncollinear. Physically, a coupling of the type (15) corresponds to either of the following three cases: (i) A noncollinear ground state, such as the helical or canted spin configurations observed in many rare-earth compounds, the itinerant helical spin-density wave in $fcc$ iron, or domain walls in ferromagnets. (ii) Excitations out of a collinear ground state, such a magnons and solitons in a ferro or antiferromagnet. (iii) A collinear state with the quantization axis chosen to be different from the axis of polarization (e.g., a ferromagnet polarized along the $x$-axis, but with $z$ chosen as the spin quantization axis). For the present purpose, of developing and testing a DFT for the 1DHM, case (i) is the most interesting one.

In this context the *staggered density*

$$\rho_{s,ij} = \langle \hat{\rho}_{s,ij} \rangle = \langle c_{i\uparrow}^\dagger c_{j\downarrow} \rangle \tag{18}$$

is most conveniently treated by considering it a new fundamental variable, conjugate to the externally applied field $S$, which enters the formalism on par with the particle density $n_i = \sum_\sigma \langle c_{i\sigma}^\dagger c_{i\sigma} \rangle$. An *ab initio* DFT based on the corresponding continuous variable

$$\rho_s(\mathbf{r}, \mathbf{r}') = \langle \Psi_\uparrow^\dagger(\mathbf{r}) \Psi_\downarrow(\mathbf{r}') \rangle, \tag{19}$$

where $\Psi_\sigma(\mathbf{r})$ is a field operator, has been proposed in Ref. [48]. Applications to the Overhauser spin-density wave (SDW) in one and three dimensional electron gases were reported in Ref. [49]. A first application to the 1DHM was presented in Ref. [50], where the stability of the 1DHM to small external staggered fields was investigated.

In this formalism the $xc$ energy becomes a functional of both densities, $n$ and $\rho_s$. Here we approximate this functional as

$$E_{xc}[n, \rho_s] \approx E_{xc}^{BALDA}[n] - \alpha U \sum_i |\rho_{i,i}|^2, \qquad (20)$$

where the first term is the BA-LDA described above, and the second is the $U\alpha$-approximation discussed in Refs. [48, 49, 51]. The coefficient $\alpha$ is an adjustable parameter in the same spirit as in the $X\alpha$ approximation of *ab initio* DFT. Here, however, it does not multiply the usual exchange energy (which is already taken into account by $E_{xc}^{BALDA}$), but rather the staggered Hartree term that constitutes the driving mechanism for the Overhauser SDW transition in the electron gas. The physical significance of this staggered Hartree term and the coefficient $\alpha$ has been discussed at length in Ref. [49].

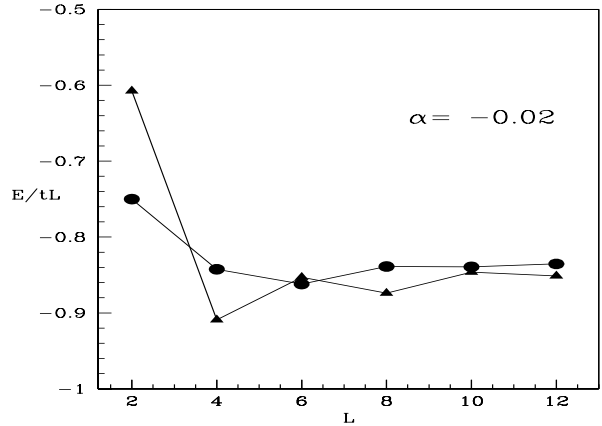

*Figure 8.* Total energy of a 1DHM with $U = 5$ and $n = 0.5$ in the presence of an external staggered field of strength $S = 0.5$. Circles: numerically exact results. Triangles: BA-LDA+$U\alpha$ approximation with $\alpha = -0.02$.

In Fig. 8 we display the total ground-state energy of a 1DHM with $U = 5$ and an even number of lattice sites, subjected to an external staggered field of strength $S = 0.5$. In the presence of this field the $z$-component of total spin, $S_z$ is not conserved anymore. One less conservation law makes the exact diagonalization much more demanding in terms of memory usage and computing time, and we present only results for up to $L = 12$ sites. The Kohn-Sham calculations, on the other hand, suffer much less from the lack of conservation of $S_z$: the appearance of matrix elements connecting up with down states in the Hamiltonian matrix is more than compensated by the absence of particle-particle interactions. In principle these calculations could be carried up to hundreds of sites, but to be able to compare with the exact results we have only gone up to $L = 12$, too.

The best agreement between the exact and the approximate values is achieved for a small negative value of $\alpha$. The physical significance of this is easily understood in light of the discussion in Refs. [49, 50, 51]: The external staggered field twists the spins away from the $z$-axis, and the staggered Hartree term always lowers the energy of a noncollinear situation with respect to a collinear one. The factor $\alpha$ corrects the staggered Hartree term, approximately taking into account the correlations not included explicitly in the density functional. A negative value for $\alpha$ means that these correlations overcompensate the energy lowering due to the staggered Hartree term. The staggered density of the system in Fig. 8 is thus exclusively due to the external field $S$. In other words, the system does not want to accomodate the SDW forced onto it by the external field. This is in agreement with our earlier finding [50] that the unperturbed system (without an external staggered field) does not have an intrinsic instability towards a noncollinear SDW.

Of course, the $U\alpha$ approximation for the $\rho_s$-dependent part of the functional (20) is much less sophisticated than the BA-LDA for the $n$-dependent one. However, the quantitative agreement between exact and approximate values, and the physically reasonable sign of the optimal value of $\alpha$, show that the $U\alpha$ approximation for the $\rho_s$-dependent part provides at least a useful starting point for further improvements [49]. Concerning the $n$-dependent part of the functional, we conclude that the BA-LDA remains computationally viable also in situations in which the inhomogeneity arises in spin space, and in which some symmetries are broken.

### 3.4 Mott insulator

In this section we return to systems that are homogeneous in the bulk, but now we choose the band filling $n$ to be exactly one. For $n = 1$ and $U \neq 0$ the 1DHM is a Mott insulator, i.e., an insulator whose gap arises from correlations, and not as a consequence of the underlying periodic lattice and the resulting single-particle band structure [6, 7, 8]. In DFT one can write the exact many-body gap as a sum of two contributions, $\Delta = \Delta_{KS} + \Delta_{xc}$, where $\Delta_{KS}$ is the difference between the highest occupied and the lowest unoccupied single-particle energies, and $\Delta_{xc}$ is defined as the discontinuity of the $xc$ potential as a function of the total particle number [52, 53, 54]

$$\Delta_{xc} = \left.\frac{\delta E_{xc}[n]}{\delta n}\right|_{N+\delta} - \left.\frac{\delta E_{xc}[n]}{\delta n}\right|_{N-\delta}, \qquad (21)$$

where $\delta \to 0^+$. The DFT characterization of a pure band insulator is $\Delta = \Delta_{KS}$, while that of a pure Mott insulator is $\Delta = \Delta_{xc}$. In general, of course, both contributions are present simultaneously. Two questions then immediately pose themselves: (1) which of the two is the dominating contribution in a given system, and (2) which of the two is reproduced by common approximate functionals?

As it turns out, the usual LDA and common gradient-corrected functionals do not have any discontinuity, and thus always predict $\Delta_{xc} = 0$. This is the origin of the so-called band-gap problem of DFT. On the other hand, the BA-LDA naturally has a discontinuity at $N = L$, where the underlying homogeneous 1DHM undergoes its metal-insulator transition [31]. Recently we have systematically investigated the resulting $\Delta_{xc}$ [15]. In fact, it is simple to calculate $\Delta_{xc}$ explicitly in the thermodynamic limit of a homogeneous system, by substituting our expression (14) with (8), (9) and (10) into Eq. (21). The result is precisely our Eq. (11) for the total gap, which was earlier calculated from the difference of ionization energy and electron affinity $I - A$. These latter two quantities can be obtained from ground-state energies according to

$$I = E(N-1) - E(N) \qquad (22)$$
$$A = E(N) - E(N+1). \qquad (23)$$

The agreement between both expressions for $\Delta$ does not come as a surprise, because in the thermodynamic limit of a homogeneous 1DHM $\Delta_{KS} \equiv 0$ (the single-particle gap vanishes and a band-structure calculation would predict the system to be a metal), so that $\Delta \equiv \Delta_{xc}$. The fact that we recover this equality shows that our expression for $e_{xc}(n, t, U)$ is reliable enough that both ways of calculating the gap from it lead to the same result.

The importance of the discontinuity in the $xc$ functional is illustrated in Fig. 9, which displays the size of the gap obtained numerically from the BA-LDA [15]; from the continuous pseudo LDA of Refs. [20, 21] (GS-LDA); directly from Eq. (11), which also follows from the BA-LDA, but is valid only for $L \to \infty$; and, for $L = 10$, also from exact diagonalization. Two conclusions can be drawn immediately from these data: (i) The exact gap is well approximated by the BA-LDA gap, but widely underestimated by the pseudo-LDA gap. This is due to the fact that the BA-LDA gap contains an $xc$ contribution due to its discontinuity, in addition to the single-particle gap. (ii) For $L = 150$ sites the asymptotic formula (11) and the numerically determined BA-LDA gap agree already quite well, whereas the pseudo-LDA gap goes to zero as the system size is increased. This latter behaviour is easy to understand: for a functional without a discontinuity, the gap is entirely due to the difference between Kohn-Sham eigenvalues. This difference, i.e., the single-particle gap, is zero in the infinite system, and the Mott metal-insulator transition is exclusively driven by the discontinuity. A more detailed DFT analysis of the Mott gap in the 1DHM can be found in Ref. [15] and, using a somewhat different approach, Ref. [22].

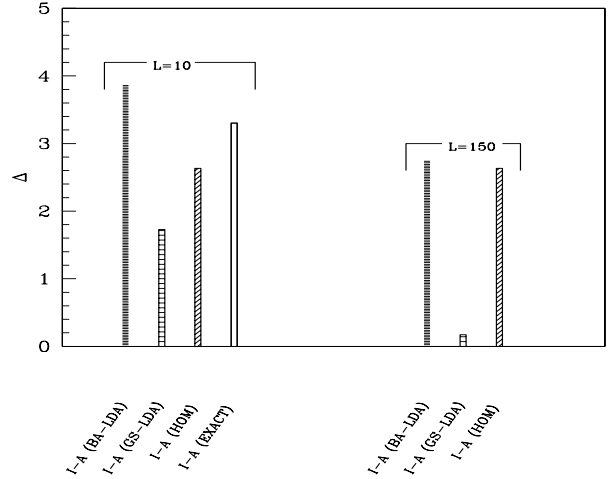

*Figure 9.* Energy gap of the 1DHM with $U = 6$, calculated for two different lattice sizes, with the methods indicated on the bars.

The intrinsic discontinuity of the BA-LDA is of course a very desirable feature for the calculation of energy gaps. The *ab initio* LDA is based on a charged Fermi liquid (a perfect metal), and the local approximation of the $xc$ energy of the inhomogeneous system by that of this homogeneous reference system amounts to locally treating the inhomogeneous system as a metal — even when it is not. Conversely, the BA-LDA is bound to locally treat a metallic system as an insulator if the local occupation number is equal to 1 within the precison of the calculation. In both cases the problem arises because the local density (at one site only) is not enough to tell whether the system should be a metal or an insulator. The resulting *inverse band-gap problem* of the BA-LDA manifests itself as a possible lack of self-consistent metallic solutions when one of the local occupation numbers comes to within $5 \times 10^{-3}$ to 1. For this reason there are some data points missing around $n = 1$ in Fig. 7. Luckily, such situations are rare: inhomogeneous Luttinger liquids with local occupation numbers very close to those for which the corresponding homogeneous system becomes a Mott insulator are realized only for some small regions in parameter space. Nevertheless, attempts to improve the BA-LDA functional in these regions, e.g., by smoothing out the discontinuity or by employing self-interaction corrections, are currently being made.

## 4.  Summary and outlook

Density-functional theory provides a way to couch the many-body problem in terms of intensive density-like variables instead of wave functions (the Hohenberg-Kohn theorem), and a practical means of extracting observables from effective single-particle equations (the Kohn-Sham scheme). Both of these achievements have had a large impact on *ab initio* calculations, but the utility of DFT extends beyond any particular Hamiltonian: Many model Hamiltonians (in particular all that are expressed in terms of intensive density-like variables) can also be analysed with DFT. The formal proof of a Hohenberg-Kohn theorem and the setting up of a Kohn-Sham scheme are straightforward transcriptions of the corresponding *ab initio* procedures.

By contrast, the construction of suitable $xc$ functionals is a more subtle matter, and crucially depends on the particular physics incorporated into the chosen model Hamiltonian [14]. In this work we have employed the Bethe Ansatz to develop an LDA-type functional for the one-dimensional Hubbard model. The resulting BA-LDA provides a very convenient and surprisingly accurate approach to large and inhomogeneous systems.

Clearly, the present work only represents a beginning. Some ways in which our functional can be improved have been mentioned at the end of Sec. 2 and of Sec. 3.4, respectively. Interesting applications of such functionals include the study of Friedel oscillations, determination of anomalous exponents, the calculation of finite-size effects, investigation of spin-density, charge-density, and bond-order waves [55], and the study of superlattices in the 1DHM. More generally, the extension of the present work to other model Hamiltonians (which require a different approach to the construction of functionals) will provide fertil ground for further applications of DFT. Work on the Heisenberg model is in progress [56].

The combination of computational efficiency with reasonable accuracy and applicability to large and inhomogeneous systems is the reason for the popularity of DFT in *ab initio* calculations in condensed-matter physics and quantum chemistry. Here we find that BA-based DFT for the 1DHM displays the same combination of features. We have thus reasons to hope that the BA-LDA may find fruitful applications in future studies of the 1DHM and other model Hamiltonians. Conversely, insights gained from such model calculations may also turn out to be useful for further development of *ab initio* DFT. The above findings about the energy gap in low-dimensional systems are one example [15], and the prospect of one day constructing an *ab initio* LDA for Luttinger liquids may be another.

## Acknowledgments

This work was supported by FAPESP, CAPES, and CNPq.